\magnification=\magstep1
\def\ni{\noindent}
\def\spose#1{\hbox to 0pt{#1\hss}}
\def\gta{\mathrel{\spose{\lower 3pt\hbox{$\mathchar"218$}}
     \raise 2.0pt\hbox{$\mathchar"13E$}}}
\def\lta{\mathrel{\spose{\lower 3pt\hbox{$\mathchar"218$}}
     \raise 2.0pt\hbox{$\mathchar"13C$}}}

\centerline{\bf Gamma-ray Burst Energetics}

\bigskip
\centerline{\bf Pawan Kumar}
\medskip
\centerline{Institute for Advanced Study, Princeton, NJ 08540}

\vskip 1.5cm

\centerline{\bf Abstract}
\bigskip\medskip

\baselineskip = 14pt

We estimate the fraction of the total energy in a Gamma-Ray Burst (GRB) 
that is radiated in photons during the main burst. Random internal
collisions among different shells limit the efficiency for converting
bulk kinetic energy to photons. About 1\% of the energy of explosion 
is converted to radiation, in 10--10$^3$ kev energy band in the observer 
frame, for long duration bursts (lasting 10s or more);
the efficiency is significantly smaller for shorter duration bursts.
Moreover, about 50\% of the energy of the initial explosion could be 
lost to neutrinos during the early phase of the burst if the initial 
fireball temperature is $\sim$ 10 Mev. If isotropic, the total energy 
budget of the brightest GRBs is $\gta 10^{55}$erg, a factor of $\gta 20$ 
larger than previously estimated.
Anisotropy of explosion, as evidenced in two GRBs, could reduce
the energy requirement by a factor of 10-100. Putting these two effects
together we find that the energy release in the most energetic bursts
is about 10$^{54}$ erg.

\bigskip
\hskip0.3cm{\it Subject headings:\rm~ gamma-rays: bursts -- gamma-rays: theory}

\vfill\eject

\centerline{\bf 1. Introduction}
\bigskip

The short, milli-second, time variability of gamma-ray bursts is 
believed to arise in internal shocks i.e. when faster moving ejecta 
from the explosion collides with slower moving material ejected at 
an earlier time (Paczynski \& Xu 1994, Rees \& M\'esz\'aros 1994, 
Sari \& Piran 1997). The optical identification and measurement 
of redshifts for five GRBs have determined their distances and 
the amount of energy that would be radiated in an isotropic 
explosion (eg. Metzger et al. 1997, Kulkarni et al. 1998, 
Kelson et al. 1999, Piran 1999 and references therein). 
In three of these cases (GRB 971214, 980703 and 990123), the total
isotropic energy radiated  is estimated to be in excess of 10$^{53}$ erg. 
For GRB 990123 the isotropic energy in the gamma-ray burst is estimated 
to be 3.4x10$^{54}$ erg. However, the steepening of the fall-off of the 
optical light curve, $\sim 2$ days after the explosion, suggests that 
the explosion was not isotropic, and the total radiated energy might 
only be $\sim 6$x$10^{52}$ erg (Kulkarni et al. 1999; M\'esz\'aros \& 
Rees, 1999). There is little evidence for beaming in the other two cases.

The energy radiated in photons in gamma-ray bursts is only a fraction
of the total energy released in the explosion. Collisions of shells or 
ejecta from the central source, believed to produce the highly variable 
gamma-ray burst emission, converts but a small fraction of the kinetic 
energy of the ejecta into thermal energy which is shared among protons, 
electrons and magnetic field. If the initial temperature of the fireball 
is larger than a few Mev then a fraction of the fireball energy is lost 
to neutrinos. Thus a significantly larger amount of energy than 
`observed' must be released in these explosions. The purpose of this 
paper is to provide an estimate for the radiative efficiency of GRBs 
in the framework of the internal shock model (\S2). 
Some aspects of the work presented here has been previously
considered by Kobayashi et al. (1997) and Daigne \& Mochkovitch (1998).
The main points are summarized in \S3.

\bigskip
\centerline{\bf 2.  Gamma-ray burst energetics}
\bigskip

A fraction of the kinetic energy of ejecta in GRBs is converted 
into photons as a result of internal collision during the main burst.
This efficiency factor is calculated below in \S2.1.
Just after the explosion, when the adiabatic cooling is small and
the temperature of the fireball is several Mev, neutrinos are copiously
produced and carry away a fraction of the energy of the explosion.
The fraction of energy lost to neutrinos is calculated in \S2.2.

\bigskip
\centerline{\bf 2.1 Efficiency of internal shocks}
\medskip

The efficiency of conversion of the kinetic energy of ejecta to 
radiation via internal shocks has been considered by Kobayashi et 
al. (1997) and Daigne \& Mochkovitch (1998). 
There are several differences between the calculation presented 
here and previous works. 
One of which is that we calculate synchrotron emission from forward and 
reverse shocks in colliding shells and compton up scattering of photon energy, 
by solving appropriate equations for shock and radiation, to determine the 
observed fluence in the energy band 10--10$^3$ kev.
We also take into consideration that about one--third of the 
total thermal energy produced in colliding shells is taken up 
by electrons, and only this fraction is available to be radiated away.
Finally, we treat in a consistent manner 
energy radiation in shell collisions when the fireball is optically 
thick to Thomson scattering. In this case, photons do not escape the 
expanding ejecta but instead deposit their energy back into shells 
and increase the kinetic energy of ejecta. Most of this kinetic
energy is not converted back to thermal energy until some later 
time when interstellar material is shocked. The reason for this 
is that shell mergers reduce the relative Lorentz factor of
remaining shells and their subsequent mergers produce less thermal energy.
The optical depth is important for bursts of duration ten seconds or 
less (hereafter referred to as short duration bursts).

We model the central explosion as resulting in random ejection of 
discrete shells each carrying a random amount of energy ($\epsilon_i$), and
with a random Lorentz factor ($\gamma_i$). The baryonic mass of i-th shell 
($m_i$) is set by its energy ($\epsilon_i$) and  $\gamma_i$; 
$m_i = \epsilon_i/(c^2 \gamma_i)$. The time interval between the 
ejection of two consecutive shells is taken to be a random number
with mean time interval such as to give the desired total burst duration.
The Lorentz factor of shells is taken  to be uniformly distributed between
a minimum ($\gamma_{min}=5$) and a maximum ($\gamma_{max}$) value. 
The energy conversion efficiency is more or less independent of the 
number of shells ejected in the explosion so long as the number of 
shells is greater than a few. 

When two cold shells i \& j collide and merge the thermal energy produced is

$$ \Delta E = \gamma_f \left[ (m_i^2 + m_j^2 + 2 m_i  m_j 
\gamma_{r})^{1/2} - (m_i+m_j)\right]c^2, $$
where $\gamma_{r} = \gamma_i\gamma_j(1 - v_i v_j)$ is the Lorentz 
factor corresponding to the relative speed of collision, and 
$\gamma_f = (m_i\gamma_i + m_j\gamma_j) (m_i^2 + m_j^2 + 
2 m_i m_j\gamma_{r})^{-1/2}$
is the final Lorentz factor of the merged shells. The 
energy $\Delta E$ is shared among protons, electrons 
and magnetic field. In equipartition, electrons take up one third of 
the total energy, which is available to be radiated.  In collisions 
involving two equal mass shells with $\gamma_{r}=2$, 6\% of the
energy can be radiated away, whereas collisions with $\gamma_{r}=10$
result in a loss of 19\% of the energy. The average relative Lorentz
factor of shell collisions is about 2 if shells are randomly ejected
in a relativistic explosion. Thus the average bolometric radiative
efficiency of internal shocks is about 6\%. Approximately 1/4 of
the total radiative energy lies in the energy band 10--10$^3$ kev, 
and therefore the effective radiative efficiency, in the observed energy
band, of internal shocks is about 1\%. More precise results from
numerical simulations are presented below.

The time scale for the transfer of energy from protons to electrons 
due to Coulomb collisions, even when the number density of protons
$\sim 10^{13}$ cm$^{-3}$ at a time when the fireball is just becoming 
optically thin, is much longer than the dynamical time and so we 
assume that there is little transfer of energy from protons to electrons
on the time scale of interest for internal shocks.

The synchrotron cooling time, $t_s$, is typically much less than
the dynamical time within the first few minutes of the burst and does not
limit the efficiency of GRBs. In any case we include the effect of finite
synchrotron cooling time on the radiative efficiency.
We also include the inverse Compton cooling of electrons to calculate the 
spectrum and the fraction of thermal energy radiated away in internal shocks.

Following each shell collision we calculate the thermodynamical state
of the shocked gas, and the emergent photon spectrum resulting from
synchrotron emission plus the inverse compton scattering.
The optical depth for emergent photons to Thomson scattering is
calculated by following their trajectory 
along with the trajectory of shells. If the optical depth of the fireball 
is greater than a few, the photon energy gets converted back to the 
energy of bulk motion via adiabatic expansion and the momentum deposit 
by photons. The energy, $\delta E_j$, and momentum, $\delta P_j$,
incident on a shell $j$ (as measured in its rest frame), by photons 
created in a colliding shell an optical depth $\tau_j$ away, which is 
moving with a relative velocity $v_{cj}$ toward the j-th shell, is given by

$$ \delta E_j = \eta_j \gamma_{cj} (1+v_{cj})^2, $$
and
$$ \delta P_j = {\eta_j\over c(v_{cj}\gamma_{cj})^3}
    \left[ \gamma_{cj}^4 (1 + v_{cj})^2 - 4\gamma_{cj}^2 (1 + v_{cj}) v_{cj}
 + 2\ln\gamma_{cj}^2(1+v_{cj}) - 1\right], $$
where $\eta_j = \Delta E \exp(-\tau_j) [1 - \exp(-\delta\tau_j)]/(6\gamma_f)$
is the energy incident on the j-th shell if it were stationary 
with respect to the center of momentum of the colliding shells,
$\gamma_f$ is the Lorentz factor of merged shells,
and $\delta\tau_j$ is the optical depth of the j-th shell. For $\tau_j$
dominated by scattering opacity the flux from a steady source is 
attenuated by a factor of $\sim 1/\tau_j$ instead of $\exp(-\tau_j)$ given 
above. However, the energy/momentum received from a transient source on the 
short, photon transit time, is reduced by a factor of $\exp(-\tau_j)$. 
The remainder of the energy/momentum is received on a
longer time scale, of order photon diffusion time, and is included
in our numerical computation where appropriate.
For the elastic Thomson scattering by cold electrons the incident photon
energy is only partially absorbed in optically
thick shells as a result of the adiabatic expansion of the shell.
The energy--momentum intercepted by a shell which is moving away from
the energy producing shell is much smaller and is given by

$$ \delta E_j = {\eta_j \over (1+v_{cj})^2 \gamma_{cj}^3}, $$
and
$$ \delta P_j = {\eta_j\over c (v_{cj} \gamma_{cj})^3} \left[
   {(1+v_{cj})^2 - 1\over (1+v_{cj})^2} - {4 v_{cj}\over 1+v_{cj}}  
      + 2\ln(1+v_{cj})\right]. $$

\ni The energy and momentum absorbed by the shell determines the change 
to its bulk velocity and its expansion which we include
in our numerical simulation to determine the radiative efficiency
of internal collisions. Also included in our calculation is the 
conversion of the thermal energy of protons and the magnetic field 
to bulk motion as a result of adiabatic expansion.

The radiative efficiency, $\eta$, of a burst is defined as the total 
energy radiated in the energy band 10--10$^3$ kev, during a time interval 
in which shell collisions take place, divided by the total energy released 
in the explosion. 

Figure 1 shows a plot of $\eta$ as a function of burst duration.
The total energy in bursts, in all of the cases shown in the figure,
was taken to be $10^{52}$ erg, independent of the burst duration.
The value of $\eta$ is found be about 1\% for long duration bursts.
The bolometric radiative efficiency of random internal shocks is
found to be larger by a factor of about 4.
The efficiency decreases with decreasing duration 
(for a fixed $\gamma_{max}$). Internal shocks are very inefficient
for short duration bursts, because of photon trapping, as a number of shell 
collisions occur when the shell radii are small and the fireball is 
optically thick. For instance, the radiative efficiency for bursts of 1 sec
duration is about 0.2\% if $\gamma_{max}=200$.
The radiative efficiency for short duration bursts can increase
significantly if the Lorentz factor of ejecta is larger in shorter
duration bursts (see fig. 1). Choice of a different distribution function 
for the Lorentz factor of ejecta has little effect on the efficiency
of long duration bursts. However, the efficiency of short duration bursts
can increase significantly if the width of the distribution function
is taken to be small so that shells collide at larger radii enabling
photons to escape freely; For instance, in the case where $\gamma_{min} =50$
\& $\gamma_{max}=200$ the radiative efficiency is nearly constant, $\eta
\approx 0.006$, for bursts of duration 1 sec and longer (see fig. 1).
The efficiency for short duration bursts is also enhanced if they are
less energetic than longer duration bursts thereby requiring smaller
baryonic loading.

\bigskip
\centerline{\bf 2.2 Energy loss due to neutrino production}
\bigskip

Some fraction of gamma-ray bursts display variability on milli-second
time scale if not less. The energy of explosion in these cases is expected
to be generated in a region of size about 100 km.  If the total energy 
release in an explosion underlying a GRB is $E$ and it involves ejection 
of $N$ shells, each of which have an initial radius of $r_0$  then the 
mean initial temperature of shells is 
$T_0 = [3 E_n/(4Na\pi r_0^3)]^{1/4}= 20.6$ Mev $E_{53}^{1/4} r_{100}^{-3/4} 
N^{-1/4}$; where $a$ is the radiation constant, $E_{53}$ is 
energy in units of 10$^{53}$ erg, and $r_{100}=r_0/100$km. We 
note that the energy of the explosion ($E$) is greater than the observed 
energy in the gamma-ray emission by a factor of at least ten because of 
the inefficiency of photon production discussed in \S2.1. Moreover,
the value of $E$ that should be used in calculating the temperature 
is the total isotropic energy of explosion and not the reduced energy 
due to finite opening angle of jet, so long as the jet was produced in 
the initial explosion and not by some collimation effect of the
surrounding medium subsequent to a spherical explosion.
Thus $E \approx 10^{53}$ erg is a reasonable value for 
the five GRBs with known redshift distance.

Neutrinos produced by $e^-$--$e^+$ annihilation, and the decay of meuons  
and pions result in a loss of a fraction of the energy of explosions.
The energy loss rate due to $e^-$--$e^+$ annihilation is given by

$$ {d E_n\over dt} =  - 2 n_e c \sigma_e \epsilon_e  (4\pi r^2 r_0 n_e), $$
where $E_n=E/N$, $n_e$ is the number density of electrons, $\epsilon_e$ is the 
mean thermal energy of electrons, $4\pi r^2 r_0$ is the volume of 
the shell in its comoving frame when the shell has expanded to
a radius $r$ (the shell thickness, $r_0$, is very nearly constant
in the initial acceleration phase), and $\sigma_e = 
2\times 10^{-44} (\epsilon_e/1 MeV)^2$cm$^2$ is the effective 
cross section for $e^+$ and $e^-$ annihilation to produce neutrinos of
all different flavors. Since $E_n \approx 12\pi r^2 r_0 n_e\epsilon_e\gamma$,
$n_e = 2.34\times 10^{34}T_{10}^3$ cm$^{-3}$ ($T_{10} = T/10$ Mev), and 
$\epsilon_e = 3.15 k T$, we find

$$ {d \ln E\over dt} = - {9.5\times 10^3\over \gamma} T_{10}^5. $$

Initially the Lorentz factor of shells ($\gamma$) increases linearly with 
their radius and the temperature declines as the inverse of the radius.
Using these relations we can integrate the above equation and find that

$$  \ln\left[{E(2 t_0)\over E(t_0)}\right] = -{1.9\times 10^3} 
t_0 \left({T_0\over 10{\rm Mev}}\right)^5, $$

\ni where $t_0$ is the larger of $r_0/c$ and the time when the shell 
becomes optically thin to neutrinos; shells become optically thin to
electron neutrinos when $T_0\le 10.2$ Mev. A neutrino propagating
outward sees the mean electron energy and density decrease and 
therefore the opacity for scattering in an expanding medium is smaller 
than a corresponding static shell. 

For $r_0 = 10^7$ cm and $T_0 = 7$ Mev we find that 10\% 
of the energy of the explosion is lost to neutrinos from
$e^+$--$e^-$ annihilation, and for $T_0 = 10$ Mev, 50\% of 
the energy is lost.

We next calculate the fraction of energy carried away by neutrinos
produced by the decay of muons and pions. Let us consider an unstable 
particle ($\mu^{\pm}$ or $\pi^{\pm}$) of mass $m_d$, that has a lifetime
of $t_d$, the number density $n_d$, and the amount of energy
carried by neutrinos when it decays is $\epsilon_\nu$. In the 
temperature range of interest to us, these particles are created
by $e^{\pm}$ interaction on time scale short compared to their 
decay time and so their number density is given by the thermal
distribution i.e.

$$ n_d = 10.5\, T^3 \left( {m_d\over kT}\right)^{3/2} \exp(-m_d c^2/kT)
 \;{\rm cm}^{-3}. $$

The rate of loss of energy of the explosion to escaping neutrinos 
produced by the decay of these particles is given by

$$ {d E\over dt} = - {8\pi r^2 r_0 n_d \epsilon_\nu\over t_d}\approx
   -{E\epsilon_\nu\over 8 t_d kT}
    \left({m_dc^2\over kT}\right)^{3/2} \exp(-m_d c^2/kT). $$

\ni This equation can be easily integrated to yield,\footnote{$^1$}
{The $\nu_\mu$'s produced in these decays find the shell to be
optically thin so long as the shell temperature is less than about 15 Mev.
For $T_0\gta 15$Mev the $\nu_\mu$'s are trapped in the fireball and
their distribution is thermal in equilibrium with $e^\pm$. In this
case roughly 50\% of the fireball energy is lost to neutrinos.}

$$ \ln\left[{E(2 t_0)\over E(t_0)}\right] = -{t_0\over t_d}
    {\epsilon_\nu\over 8 kT}\left({ m_d c^2\over kT_0}
    \right)^{1/2} \exp(-m_d c^2/kT). $$

For the muons $m_d = 105.66 $Mev, $t_d=2.2$x$10^{-6}$ s, 
$\epsilon_\nu\approx 70$ Mev. Thus the fraction of energy lost by 
the decay of  $\mu^{\pm}$ for $T_0 = 10$ Mev and 
$t_0=3.3$x10$^{-4}$s, is 0.5\%, whereas at $T_0=15$ Mev, 10\%
of the energy of the fireball is lost to neutrinos from muon decay.

For pions $m_d = 139.6$Mev, $t_d=2.55$x$10^{-8}$ s, 
$\epsilon_\nu\approx 29$ Mev. The fraction of energy lost by the 
decay of  of  $\pi^{\pm}$ if we take $T_0 = 10$ Mev and
$t_0=3.3$x10$^{-4}$ s is 2\%, whereas at $T_0=15$ Mev, 50\% of 
the energy of the explosion is lost to neutrinos from pion decay.

In summary, we find that a fraction of the energy of the explosion
is lost to neutrinos. The fraction lost depends on the initial
temperature of the fireball, and for plausible burst parameters
roughly half the energy of the explosion is carried away by neutrinos.
Since the typical energy of these neutrinos is about 10-30 Mev they
are undetectable from a typical GRB source at $z\sim 1$. The total
energy in high energy neutrinos, $\epsilon_\nu\gta 10^{14}$ev, 
produced in internal shocks, is about two orders of magnitude smaller
than the energy in the 1--30 Mev neutrinos considered here. However,
the much larger cross-section for the high energy neutrinos  
makes them accessible to the large neutrino detections under 
construction (Waxman and bahcall, 1998).

\bigskip
\centerline{\bf Summary and discussion}
\medskip
We find that the efficiency for internal shocks to convert the
energy of explosion to radiation in the energy band 10---10$^3$ kev
is of order 1\% 
if electrons are in equipartition with protons and magnetic field. 
The efficiency is smaller if the electron energy is less than the 
equipartition value as suggested by analysis of afterglow emission 
(eg. Waxman 1997). Energy loss due to neutrino production at initial 
times, when the fireball temperature is $\sim 10$ Mev for short duration
bursts, could be significant, further reducing the energy 
available for radiation by a factor of $\sim$ two.
The bolometric radiative efficiency of random internal shocks
is found to be a factor of about 4 larger.
A recent work of Panaitescu, Spada and M\'esz\'aros (1999)
finds the radiative efficiency of internal shocks in the 50--300 kev
band to be about 1\%, and is consistent with our result.

For GRB 971214, 980703 and 990123, the total isotropic energy radiated, in 
the BATSE energy band, has been estimated from their observed 
redshifts and fluences and found to be 3x10$^{53}$, 2x10$^{53}$
and 3.5x10$^{54}$ erg respectively. The flux in higher energy
photons could increase the total energy budget by a factor of $\sim 2$.
These three bursts are the most energetic
of the five bursts for which redshifts (or lower limits to $z$)
are known. These energies should of course be corrected for beaming
and the efficiency for photon production.

It has been suggested that the energy for GRB 990123, 
3.5x10$^{54}$erg for isotropic explosion, is reduced by
a factor of about 50 due to finite beaming angle (Kulkarni et al. 1999;
M\'esz\'aros \& Rees 1999). However, the inefficiency of producing 
radiation raises the energy budget by a factor of about 100, so 
the energy in the explosion is more than 10$^{54}$ erg even if 
beaming is as large as suggested. For GRB 980703 (at $z=0.966$), 
for which there is no evidence for beaming, the energy in the 
explosion is also of order 10$^{54}$ erg. So it appears that the 
total energy of explosion for the most energetic bursts is close 
to or possibly greater than 10$^{54}$ erg. This energy is greater 
than what one can realistically hope to extract from a neutron star 
mass object.

The efficiency for gamma-ray production is significantly increased if 
photons during the main burst are produced in both internal and external
shocks.  However, since it is very difficult to get short time
variability in external shocks (Sari \& Piran, 1997) only a
small fraction of energy in highly variable bursts can arise in
external shocks. The energy requirement is also reduced if shells
ejected in explosions are highly inhomogeneous. This will be 
discussed in a future paper.

\medskip
\ni{\bf Acknowledgments:} I thank John Bahcall for encouraging me
to writeup this work and for his comments. I am grateful to
Plamen Krastev for providing accurate neutrino cross-sections, and I am
indebted to Ramesh Narayan, Tsvi Piran and Peter M\'esz\'aros for
many helpful discussions. I thank an anonymous referee for suggestions
to clarify some points in the paper.

\vfill\eject

\centerline{\bf REFERENCES}
\bigskip

\ni Daigne, F. and Mochkovitch, R. 1998, MNRAS

\ni Kabayashi, S., Piran, T. \& Sari, R. 1997, ApJ 490, 92

\ni Kelson et al., 1999, IAUC 7096

\ni Kulkarni, S.R., et al. 1998, Nature 393, 35

\ni Kulkarni, S.R. et al., 1999, astro-ph/9902272

\ni Metzger, M.R., et al. 1997, Nature, 387, 879

\ni M\'esz\'aros, P., and rees, M.J. 1999, astro-ph/9902367

\ni Narayan, R., Piran, T., and Shami, A., 1991, ApJ 379, L17

\ni Pacznski, B. and Xu, G. 1994 ApJ 427, 708

\ni Panaitescu, A., Spada, M. \& M\'esz\'aros, P., 1999, astro-ph/9905026

\ni Piran, T., 1999, to appear in Physics Reports

\ni Rees, M.J. and M\'esz\'aros, P., 1994, ApJ 430, L93

\ni Sari, R. and Piran, T., 1997, MNRAS 287, 110

\ni Waxman, E., 1997, ApJ 489, L33

\ni Waxman, E. and Bahcall,J.N., 1998, astro-ph/9807282

\vfill\eject
\centerline{\bf Figure Captions}
\bigskip

\ni Figure 1.--- The efficiency for the conversion of the 
energy of explosion to radiation, in the energy band 10--10$^3$ kev,
via internal shocks ($\eta$*100)
is shown as a function of the time duration of GRBs. The energy lost
to neutrinos is highly temperature dependent and has not been included 
in this calculation. The continuous curve corresponds to the maximum
Lorentz factor of the ejected shells to be 200, and for the dotted curve
the maximum Lorentz factor is 500. The minimum value of the Lorentz factor
in both these cases was taken to be 5. The minimum Lorentz factor for the
dashed curve was taken to be 50 \& $\gamma_{max} = 200$. 
Each point on the curve was calculated by averaging 250 realizations of 
'explosions' in which 50 shells were randomly expelled as described 
in section 2.1. The total energy in each of the explosion was taken to be
10$^{52}$ erg which was independent of the burst duration.
The radiative efficiency is almost independent of the number of 
shells ejected so long as the number is larger than a few.

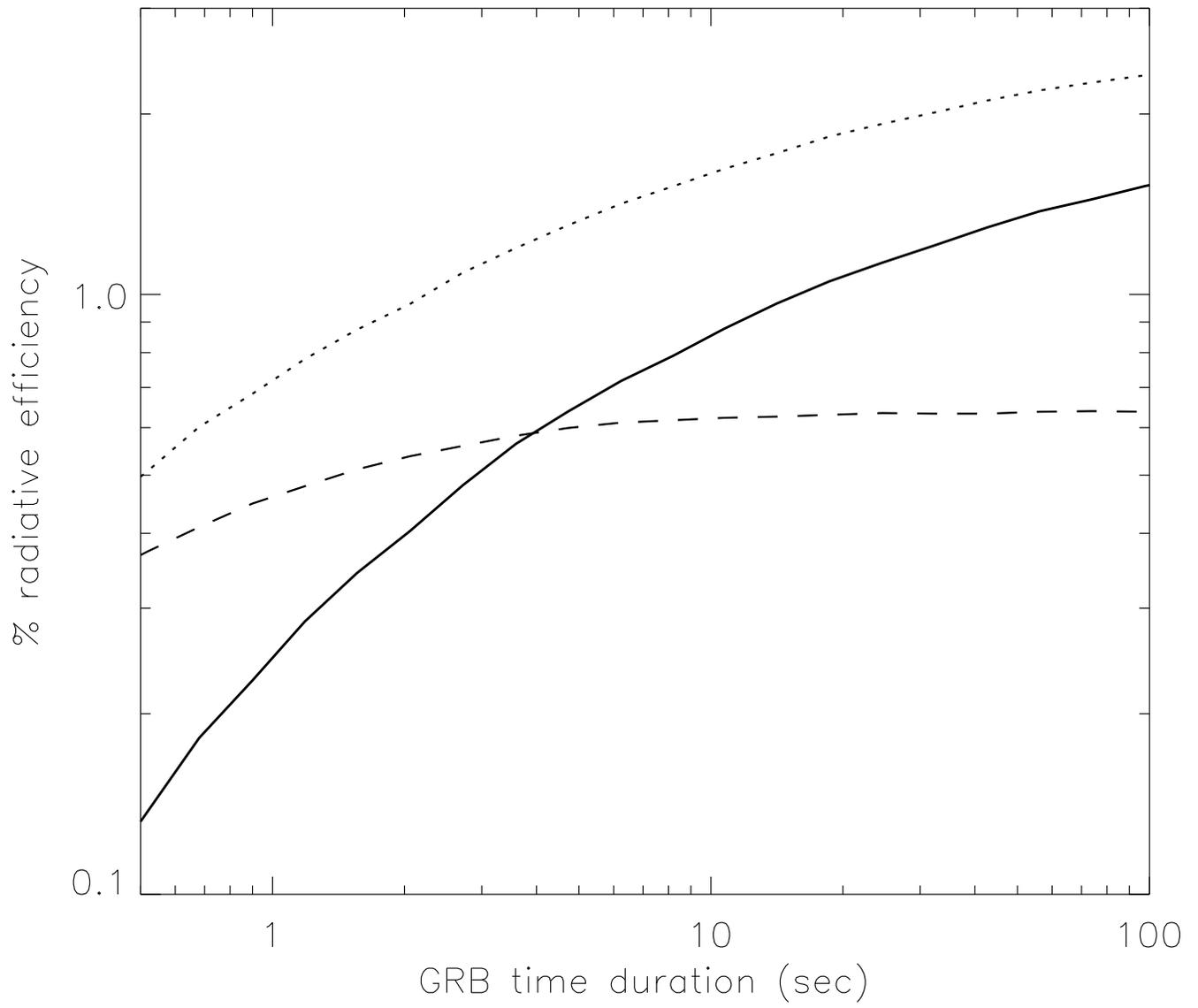
\bye